\documentclass{article}
\usepackage{spconf,amsmath,graphicx}
\usepackage{stmaryrd}
\usepackage{bm}
\usepackage{amssymb}
\usepackage{booktabs}
\usepackage{graphicx}
\usepackage{cite}
\usepackage{float}
\usepackage{graphicx}
\usepackage{stfloats}
\usepackage{amsmath}
\usepackage{epstopdf}
\usepackage{algorithm,algorithmic}
\usepackage{multirow}
\usepackage{framed}
\usepackage[colorlinks,linkcolor=red,anchorcolor=green,citecolor=green]{hyperref}
\usepackage[numbers,sort&compress]{natbib}
\usepackage{amsthm}
\usepackage{makecell}

\newcommand{\argmin}{\arg\!\min}

\newcommand{\Cmat}{{\bf C}}
\newcommand{\Dmat}{{\bf D}}

\newcommand{\Gmat}[0]{{{\bf G}}}

\newcommand{\Wmat}[0]{{{\bf W}}}
\newcommand{\Xmat}{{\bf X}}
\newcommand{\Ymat}[0]{{{\bf Y}}}

\newcommand{\gv}[0]{{\boldsymbol{g}}}

\newcommand{\pv}[0]{{\boldsymbol{p}}}

\newcommand{\uv}[0]{{\boldsymbol{u}}}
\newcommand{\vv}{\boldsymbol{v}}
\newcommand{\wv}{\boldsymbol{w}}

\newcommand{\xv}{\boldsymbol{x}}
\newcommand{\yv}{\boldsymbol{y}}

\newcommand{\zv}{\boldsymbol{z}}

\newcommand{\Phimat}{\boldsymbol{\Phi}}

\newcommand{\thetav}{\boldsymbol{\theta}}

\newcommand{\ts}{^{\top}}
\newcommand{\ie}{{\em i.e.}}
\newcommand{\eg}{{\em e.g.}}
\newcommand{\inv}{^{-1}}

\title{Various Total Variation for Snapshot Video Compressive Imaging}
\name{Xin Yuan}
\address{Bell Labs, 600 Mountain Aveue, Murray Hill, NJ 07974, USA}
\begin{document}
%
\maketitle
\begin{abstract} 
Sampling high-dimensional images is challenging due to limited availability of sensors; scanning is usually necessary in these cases. To mitigate this challenge, snapshot compressive imaging (SCI) was proposed to capture the high-dimensional (usually 3D) images using a 2D sensor (detector). Via novel optical design, the {\em measurement} captured by the sensor is an encoded image of multiple frames of the 3D desired signal. Following this, reconstruction algorithms are employed to retrieve the high-dimensional data. Though various algorithms have been proposed, the total variation (TV) based method is still the most efficient one due to a good trade-off between computational time and performance. This paper aims to answer the question of which TV penalty (anisotropic TV, isotropic TV and vectorized TV) works best for video SCI reconstruction? Various TV denoising and projection algorithms are developed and tested for video SCI reconstruction on both simulation and real datasets.
\end{abstract}
\section{Introduction}
\label{sec:intro}
\begin{figure}[thbp!]	
	\centering
	{\includegraphics[width= 0.48\textwidth]{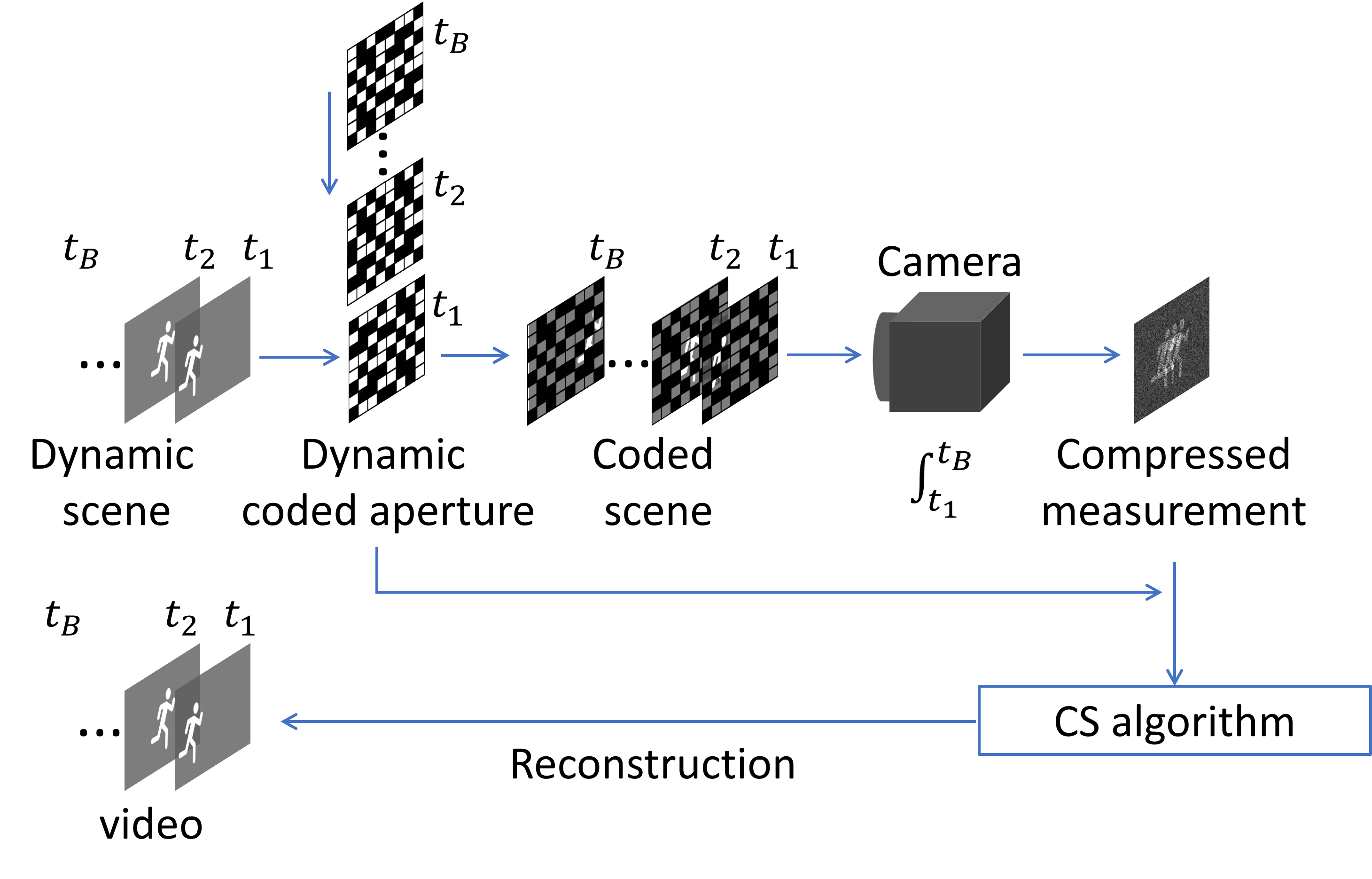}}
	\vspace{-8mm}
	\caption{Principle of snapshot video compressive sensing}
	\label{fig:principle}
	\vspace{-5mm}
\end{figure}

Snapshot compressive imaging (SCI)~\cite{Liu18TPAMI} refers to compressive imaging systems where multiple frames are mapped into a single measurement,
with video SCI~\cite{Hitomi11ICCV,Reddy11CVPR,Patrick13OE,Yuan14CVPR,Sun16OE,Sun17OE,Yuan17AO,Yuan&Pang16_ICIP,Yuan13ICIP} and spectral SCI~\cite{Gehm07,Wagadarikar08CASSI,Wagadarikar09CASSI,Yuan15JSTSP,Cao16SPM} as two representative applications. 
In video SCI shown in Fig.~\ref{fig:principle}, high-speed frames are modulated at a higher frequency than the capture rate of the camera; in this manner, each captured measurement frame can recover a number of high-speed frames, which is dependent on the coding strategy, \eg, 148 frames reconstructed from a snapshot in~\cite{Patrick13OE}. 
In spectral SCI, the wavelength dependent coding is implemented by a coded aperture (physical mask) and a disperser~\cite{Wagadarikar08CASSI,Wagadarikar09CASSI}; more than 30 hyperspectral images have been reconstructed from a snapshot measurement.  
Though it is fair to say that SCI was inspired by compressive sensing (CS)~\cite{Donoho06ITT,Candes06ITT}, the theory of SCI has just been developed in~\cite{Jalali19TIT} due to the special structure of the sensing matrix. 

Mathematically, the measurement in the SCI systems can be  modeled by
\begin{equation} \label{Eq:yPhix}
{\yv = \Phimat \xv + \gv}\,,
\end{equation}
where $\Phimat \in {\mathbb R}^{n\times nB}$ is the sensing matrix, $\xv\in {\mathbb R}^{nB}$ is the desired signal, and $\gv\in {\mathbb R}^{n}$ denotes the noise.
Unlike traditional CS, the sensing matrix considered here is not a dense matrix. 
In SCI, {\eg, video CS as in CACTI~\cite{Patrick13OE,Yuan14CVPR}}, the matrix $\Phimat$ has a very specific structure and can be written as
\begin{equation} \label{Eq:Hmat_strucutre}
{\textstyle \Phimat = \left[\Dmat_1,\dots, \Dmat_B\right]}\,, 
\end{equation}
where $\{\Dmat_k\}_{k=1}^B$ are diagonal matrices. 

As in Fig.~\ref{fig:principle}, consider that $B$ high-speed frames $\{\Xmat_k\}_{k=1}^B \in {\mathbb R}^{n_x \times n_y}$ (at timestamp $t_1,\dots, t_B$) are modulated by the masks $\{\Cmat_k\}_{k=1}^B \in {\mathbb R}^{n_x \times n_y}$, correspondingly. The 2D measurement $\Ymat \in {\mathbb R}^{n_x \times n_y}$ captured by the camera is given by
\begin{equation} \label{Eq:YXC}
{\textstyle \Ymat = \sum_{k=1}^B \Xmat_k \odot  \Cmat_k + \Gmat}\,, 
\end{equation}
where $\odot$ denotes the Hadamard (element-wise) product.
For all $B$ pixels (in the $B$ frames) at position  $(i,j)$, $i = 1,\dots, n_x$; $j = 1,\dots, n_y$, they are collapsed to form one pixel in the measurement (in one shot) as
\begin{equation}
{\textstyle y_{i,j} = \sum_{k=1}^B c_{i,j,k} x_{i,j,k} + g_{i,j}}\,.
\end{equation}
By defining 
\begin{equation} \label{Eq:xv1toB}
{{\textstyle \xv = \left[
	\xv_1\ts,\dots,\xv_B\ts
	\right]\ts}}\,,
\end{equation}
where $\xv_k = {\rm vec}(\Xmat_k)$,  and $\Dmat_k = {\rm diag}({\rm vec}(\Cmat_k))$, for $ k=1,\dots, B$, we have the vector formulation of Eq.~\eqref{Eq:yPhix}, where $n = n_xn_y$.
Therefore, $\xv \in {\mathbb R}^{n_xn_yB}$, $\Phimat\in {\mathbb R}^{n_xn_y \times (n_xn_yB)}$, and the compressive sampling rate in  SCI is equal to  $1/B$, which is defined by the hardware design.
It has recently been proved that even when $B>1$, reconstruction can be achieved with overwhelming probability~\cite{Jalali18ISIT,Jalali19TIT}.

The following task for the algorithm is to reconstruct the desired signal $\xv$ given the measurement $\yv$ and the special sensing matrix $\Phimat$ determined by the physical masks $\{\Cmat_k\}_{k=1}^B$.

\section{Solve SCI by Total Variation Regularization}
Obviously, Eq.~\eqref{Eq:yPhix} is an ill-posed problem and a regularizer is usually utilized to confine the solution. In this paper, we focus on the total variation (TV) regularization and thus solve the following problem,
\begin{equation}
\hat{\xv} = \argmin_{\xv} \frac{1}{2}\|\yv-\Phimat \xv\|_2^2 + {\rm TV}(\xv),
\end{equation}
where ${\rm TV}(~)$ denotes the TV regularizer.
Since $\xv$ inherently is a 3D data-cube in SCI, various TV can be used. For example, the {\em Anisotropic TV} (ATV) and the {\em Isotropic TV} (ITV) and moreover the TV can be imposed on each 2D frame   of the video or on the entire 3D cube.

For the ease of notation, in the following, we first define the operators:
\begin{eqnarray}
{\cal D}_h \xv_k = \Xmat_k \Dmat_h\ts, && {\cal D}_v \xv_k = \Dmat_v \Xmat_k,
\end{eqnarray}
where $\{\Dmat_h \in {\mathbb R}^{(n_y-1)\times n_y},  \Dmat_v \in {\mathbb R}^{(n_x-1)\times n_x}\}$
as the gradient operator to perform differentiation on the desired frame horizontally and vertically, respectively.

\subsection{Different TV Formulations \label{Sec:TV}}
Different TVs can thus be summarized as follows:
\begin{list}{\labelitemi}{\leftmargin=12pt \topsep=0pt \parsep=-3pt}
	\item ATV:
	\begin{equation}
		\textstyle{\rm ATV}(\xv) = \sum_{k=1}^B \left(\|{\cal D}_h \xv_k\|_1 + \|{\cal D}_v \xv_k\|_1\right).  \label{Eq:ATV_full}
	\end{equation}
	Note that the formulation of ATV2D is the same as ATV3D.
	\item ITV2D:
	\begin{equation}
		\textstyle{\rm ITV2D}(\xv) = \sum_{k=1}^B\sqrt{\|{\cal D}_h \xv_k\|_2^2 + \|{\cal D}_v \xv_k\|_2^2}. \label{Eq:ITV2D_full}
	\end{equation}
	\item ITV3D:
	\begin{equation}
	\textstyle	{\rm ITV3D}(\xv) = \sqrt{\sum_{k=1}^B \left(\|{\cal D}_h \xv_k\|_2^2 + \|{\cal D}_v \xv_k\|_2^2\right)}. \label{Eq:ITV3D_full}
	\end{equation}
\end{list}
We thus have the following problems to solve the SCI reconstruction using various TV:
\begin{list}{\labelitemi}{\leftmargin=12pt \topsep=0pt \parsep=-3pt}
	\item [1)] ATV: 
	\begin{equation} \label{Eq:SCI_ATV}
	\textstyle\hat\xv = \argmin_{\xv}\frac{1}{2}   \|\yv- \Phimat\xv\|_2^2 + \lambda \sum_{k=1}^B \text{ATV2D}(\xv_k).
	\end{equation}
	\item [2)] ITV2D:
	\begin{equation}\label{Eq:SCI_ITV2D}
	\textstyle\hat\xv = \argmin_{\xv}\frac{1}{2}   \|\yv- \Phimat\xv\|_2^2 + \lambda \sum_{k=1}^B \text{ITV2D}(\xv_k).
	\end{equation}
	\item [3)] ITV3D:
	\begin{equation} \label{Eq:SCI_ITV3D}
	\textstyle\hat\xv = \argmin_{\xv}\frac{1}{2}   \|\yv- \Phimat\xv\|_2^2 + \lambda\text{ITV3D}(\xv).
	\end{equation}
\end{list}

\subsection{Different Solvers}
The previous section have presented different TV norms and here we present different popular solvers (we are not seeking for a thorough survey here) in the literature.
The SCI reconstruction problem in Eq.~\eqref{Eq:SCI_ATV}-Eq.\eqref{Eq:SCI_ITV3D} can be solved using different frameworks.
\begin{list}{\labelitemi}{\leftmargin=12pt \topsep=0pt \parsep=-3pt}
	\item FISTA~\cite{Beck09IST}: It consists the following steps
	\begin{align}
	\zv^{(t)} &= \textstyle\thetav^{(t)} + \frac{1}{L(f)}\Phimat\ts(\yv- \Phimat \thetav^{(t)}),\\
	\xv^{(t)} &= {\rm TVdenoise}(\zv^{(t)} ), \\
	\tau^{(t+1)}&= \textstyle\frac{1+ \sqrt{1+4(\tau^{(t)})^2}}{2},\\
	\thetav^{(t+1)}&= \textstyle\xv^{(t)} + \frac{\tau^{(t)}-1}{\tau^{(t+1)}}(\xv^{(t+1)} - \xv^{(t)}),
	\end{align}
	where $\tau^{(1)}=1$ is introduced in FISTA and various TV norms in Sec.~\ref{Sec:TV} (with solutions in Sec.~\ref{Sec:TV_sol}) can be used.  $f(\xv) = \frac{1}{2}\|\yv-\Phimat\xv\|_2^2$ is continuously differentiable with Lipschitz continuous gradient $L(f)$ (derived elsewhere).
	\item TwIST~\cite{Bioucas-Dias2007TwIST}: It consists the following steps
	\begin{align}
	\zv^{(t)} &= \xv^{(t)} + \Phimat\ts(\yv- \Phimat \xv^{(t)}),\\
	\thetav^{(t)} &= {\rm TVdenoise}(\zv^{(t)} ), \\
	\xv^{(t+1)}&= (1-\alpha)\xv^{(t-1)} + (\alpha-\beta) \xv^{(t)}+\beta \thetav^{(t)},
	\end{align}
	where $\{\alpha,\beta\}$ are TwIST parameters and can be determined by the eigenvalues of $\Phimat\ts\Phimat$.
	\item GAP~\cite{Liao14GAP}: It consists the following steps:
	\begin{align}
	\xv^{(t+1)} &= \thetav^{(t)} + \Phimat\ts (\Phimat \Phimat\ts)\inv (\yv - \Phimat \thetav^{(t)})\\
	\thetav^{(t+1)} &= {\rm TVdenoise}(\xv^{(t+1)} ).
	\end{align}
	\item ADMM~\cite{ADMM2011Boyd}: We derive the ADMM framework by formulating the problem as
	\begin{eqnarray}
	\hat\xv = \textstyle\argmin_{\xv}\frac{1}{2}   \|\yv- \Phimat\xv\|_2^2 + \lambda  \text{TV}(\thetav), ~~{\rm s.t.}~ \thetav = \xv.
	\end{eqnarray}
	This can be solved by the following sub-problems:
	\begin{align}
	\xv^{(t+1)}&= \textstyle\argmin_{\xv} \frac{1}{2}\|\yv - \Phimat\xv\|_2^2 + \frac{\rho}{2}\|\xv-\thetav^{(t)} + \uv^{(t)}\|_2^2, \label{Eq:ADMM_xv}\\
	\thetav^{(t+1)}&= {\rm TVdenoise}(\uv^{(t)} + \xv^{(t+1)}),\\
	\uv^{(t+1)}&= \uv^{(t)} + \xv^{(t+1)} - \thetav^{(t+1)}.
	\end{align}
	As derived in~\cite{Yuan16ICIP_GAP}, since in SCI, $\Phimat\Phimat\ts$ is a diagonal matrix, Eq.~\eqref{Eq:ADMM_xv} can be solved element-wise and thus very efficiently and when $\rho = 0$, it will degrade to GAP.
\end{list}

Note that in each framework, there is a ``TVdenoise" step and various TV priors in previous subsection can be used.
In the following, we present various solutions of different ``TVdenoise".

\subsection{Solutions of TV Denoising \label{Sec:TV_sol}}
We now present different solvers for various TV denoising.
\begin{list}{\labelitemi}{\leftmargin=12pt \topsep=0pt \parsep=-3pt}
	\item ATV:
	\begin{list}{\labelitemi}{\leftmargin=10pt \topsep=0pt \parsep=-3pt}
		\item Clip: The iterative clipping algorithm~\cite{Selesnick09Clip} was employed in GAP-TV~\cite{Yuan16ICIP_GAP}. It is was derived by the  min-max
		property and the majorization-minimization procedure and inspired by~\cite{Beck09TV,Chambolle2005TVM,Zhu2010}.
		The full algorithm is listed in Algorithm~\ref{Algo:GAP-ATV}. 
		One key step is to %
		introduce variables $\{\wv_h, \wv_v\}$, with $\{|\wv_h|\le1, |\wv_v|\le1\}$. 
        \item Chambolle: in~\cite{Chambolle2004ATV,Chambolle2005TVM} (denoted as {\bf {ATV-Cham})}.
        \item FGP: (fast gradient projection) proposed in~\cite{Beck09TV} (denoted as {\bf {ATV-FGP}}) with solutions summarized in Algorithm~\ref{Algo:GAP-ATV}.
        Note we have used ${\rm max}(1, |\wv_h^{(s)} + \delta t \zv_h^{(s+1)}|)$ in the denominator of the update pf $\wv_h$ and similar for $\wv$, which is recommended in~\cite{Chambolle2005TVM}. This can also be changed to $1+ \delta t |\zv_h^{(s+1)}|$ as originally derived in~\cite{Chambolle2004ATV}. This also holds true for the following derivations on ITV2D and ITV3D.
      \end{list}
	\item ITV2D:
		\begin{list}{\labelitemi}{\leftmargin=10pt \topsep=0pt \parsep=-3pt}
		\item ITV2D-Cham: Following ATV-Cham, Let
		\begin{eqnarray}
		\tilde{\wv}_h^{(s+1)} &=& \wv_h^{(s)} + \delta t \zv_h^{(s+1)},\\
		\tilde{\wv}_v^{(s+1)}  &=& \wv_v^{(s)} + \delta t \zv_v^{(s+1)}.
		\end{eqnarray}
		Recall that $\tilde{\wv}_h^{(s+1)} $ can be a 3D video and we reshape it to $\tilde{\Wmat}_h\in {\mathbb R}^{n_x\times n_y \times B}$ by ignoring the boundary effects (and also dropping the index $(s+1)$), and similar to $\tilde{\Wmat}_v$. We further let $[\tilde{\Wmat}_h]_{i,j,k}$ denotes the $(i,j)$-th pixel or voxel in $k$-th frame and similar for $[\tilde{\Wmat}_v]_{i,j,k}$. We now have the update equations for 	$[{\Wmat}_h]_{i,j,k}$ and $[{\Wmat}_v]_{i,j,k}$, which correspond to $\wv_h$ and $\wv_v$, respectively.
		\begin{align}
		[{\Wmat}_h]_{i,j,k} &=  \textstyle\frac{[\tilde{\Wmat}_h]_{i,j,k}}{{\rm max}\left(1, \sqrt{[\tilde{\Wmat}_h]_{i,j,k}^2 + [\tilde{\Wmat}_v]^2_{i,j,k}}\right)}, \label{Eq:whijk_ITV2D}\\
		[\Wmat_v]_{i,j,k} &= \textstyle \frac{[\tilde{\Wmat}_v]_{i,j,k}}{{\rm max}\left(1, \sqrt{[\tilde{\Wmat}_h]_{i,j,k}^2 + [\tilde{\Wmat}_v]^2_{i,j,k}}\right)}. \label{Eq:wvijk_ITV2D}
		\end{align}
		\item ITV2D-FGP:
		Similar to ITV2D-Cham, we only need to change the update equations of $\pv_h$ and $\pv_v$ in ATV-FGP.
	\end{list}
	\item [3)] ITV3D:
	The ITV3D denosing step can be solved by the the algorithm proposed in~\cite{Chan_2008dualTV} (denoted as {\bf {ITV3D-VTV}}), or the FGP (fast gradient projection) proposed in~\cite{Beck09TV} (denoted as {\bf {ITV3D-FGP}}).
	Regarding the solution, the difference lies in Eqs.~\eqref{Eq:whijk_ITV2D}-\eqref{Eq:wvijk_ITV2D} and we now have
	\begin{align}
	[{\Wmat}_h]_{i,j,k} &=  \textstyle\frac{[\tilde{\Wmat}_h]_{i,j,k}}{{\rm max}\left(1, \sqrt{\sum_{k=1}^B\left([\tilde{\Wmat}_h]_{i,j,k}^2 + [\tilde{\Wmat}_v]^2_{i,j,k}\right)}\right)}, \label{Eq:whijk_ITV3D}\\
	[\Wmat_v]_{i,j,k} &=  \textstyle\frac{[\tilde{\Wmat}_v]_{i,j,k}}{{\rm max}\left(1, \sqrt{\sum_{k=1}^B\left([\tilde{\Wmat}_h]_{i,j,k}^2 + [\tilde{\Wmat}_v]^2_{i,j,k}\right)}\right)}. \label{Eq:wvijk_ITV3D}
	\end{align}
	Similar changes will happen for $\pv_h$ and $\pv_v$ for ITV3D-FGP.
\end{list}

\begin{algorithm}[h!]
	\caption{GAP-ATV-Clip/Cham/FGP for SCI}
	\begin{algorithmic}
		\small
		\REQUIRE Input measurements $\yv$, sensing matrix $\Phimat$.
		\STATE Initialize $\thetav^{(0)} = \Phimat\ts\yv = \vv^{(0)} , \uv^{(0)} = {\bf 0},\wv_h^{(0)} =\wv_v^{(0)}= {\bf 0},\rho, \lambda$, MaxIter and In-Iter (for TV denoising).
		\FOR{$t=0$ {\bfseries to} MaxIter }
		\STATE 	$\xv^{(t+1)} = \thetav^{(t)} + \Phimat\ts (\Phimat \Phimat\ts)\inv (\yv - \Phimat \thetav^{(t)})$.
		\STATE Select one algorithm from the following boxes.\\
		\fbox{\begin{minipage}{22em}
				{ \STATE \% ATV-Clip
					\FOR{$s=0$ {\bfseries to} In-Iter}
					\STATE  $\thetav_h^{(s+1)}= \xv^{(t+1)} -{\cal D}_h\ts \wv_h^{(s)}$,
					\STATE$\thetav_v^{(s+1)}= \xv^{(t+1)} -{\cal D}_v\ts \wv_v^{(s)}$,
					\STATE  $\wv_h^{(s+1)} ={\rm clip}\left(\wv_h^{(s)} + \frac{1}{\alpha} {\cal D}_h\thetav_h^{(s+1)},2\lambda \right)$,
					\STATE  $\wv_v^{(s+1)} ={\rm clip}\left(\wv_v^{(s)} + \frac{1}{\alpha} {\cal D}_v\thetav_v^{(s+1)},2\lambda \right).$
					\ENDFOR
					\STATE $\thetav^{(t+1)} = \thetav_h + \thetav_v - \xv^{(t+1)}$.}
		\end{minipage}}
		\fbox{\begin{minipage}{22em}
				{ \STATE \% ATV-Cham
					\STATE Initialize $\delta t = 1/8$, $\pv_d = 0$.  
					\FOR{$s=0$ {\bfseries to} In-Iter}
					\STATE  $\zv^{(s+1)} = \pv_d^{(s)} - \frac{\xv^{(t+1)}}{\lambda}$,
					\STATE   $\zv_h^{(s+1)}= {\cal D}_h\zv^{(s+1)}$,~~  $\zv_v^{(s+1)}= {\cal D}_v\zv^{(s+1)}$,
					\STATE  $\wv_h^{(s+1)} =\frac{\wv_h^{(s)} + \delta t \zv_h^{(s+1)}}{{\rm max}(1, |\wv_h^{(s)} + \delta t \zv_h^{(s+1)}|)}$,
					\STATE  $\wv_v^{(s+1)} =\frac{\wv_v^{(s)} + \delta t \zv_v^{(s+1)}}{{\rm max}(1, |\wv_v^{(s)} + \delta t \zv_v^{(s+1)}|)}$, 
					\STATE  $\pv^{(s+1)}_d =  {\cal D}_h\ts \wv_h^{(s+1)} + {\cal D}_v\ts \wv_v^{(s+1)}$.
					\ENDFOR
					\STATE $\thetav^{(t+1)} = \xv^{(t+1)} - \lambda \pv^{(s+1)}_d$.}
		\end{minipage}}\\
		\fbox{\begin{minipage}{22em}
				{ \STATE \% ATV-FGP
					\STATE Initialize $\nu^{(0)}$.  
					\FOR{$s=0$ {\bfseries to} In-Iter}
					\STATE  $\thetav^{(s+1)} = \xv^{(t+1)} - \lambda({\cal D}_h\ts \wv_h^{(s)} + {\cal D}_v\ts \wv_v^{(s)})$,
					\STATE   $\zv_h^{(s+1)}= {\cal D}_h\thetav^{(s+1)}$,~~$\zv_v^{(s+1)}= {\cal D}_v\thetav^{(s+1)}$,
					\STATE  $\pv_h^{(s+1)} =\frac{\wv_h^{(s)} + \frac{1}{8\lambda} \zv_h^{(s+1)}}{{\rm max}(1, |\wv_h^{(s)} + \frac{1}{8\lambda}  \zv_h^{(s+1)}|)}$,
					\STATE  $\pv_v^{(s+1)} =\frac{\wv_v^{(s)} + \frac{1}{8\lambda} \zv_v^{(s+1)}}{{\rm max}(1, |\wv_v^{(s)} + \frac{1}{8\lambda}  \zv_v^{(s+1)}|)}$, 
					\STATE $\nu^{(s+1)} = \frac{1+\sqrt{1+4(\nu^{(s)})^2}}{2}$,
					\STATE  $\wv^{(s+1)}_h =  \pv_h^{(s+1)}  + \frac{\nu^{(s)}-1}{\nu^{(s+1)}} (\pv_h^{(s+1)} - \pv_h^{(s)} )$,
					\STATE  $\wv^{(s+1)}_v =  \pv_v^{(s+1)}  + \frac{\nu^{(s)}-1}{\nu^{(s+1)}}(\pv_v^{(s+1)} - \pv_v^{(s)} )$.
					\ENDFOR
					\STATE $\thetav^{(t+1)} = \thetav^{(s+1)}$.}
		\end{minipage}}\\
		\ENDFOR
		\STATE {\bf Output}  $\xv$.
	\end{algorithmic}
	\label{Algo:GAP-ATV}
\end{algorithm}

\begin{table}[htbp!]
	\caption{\small Different frameworks and various TV denoising algorithms to solve SCI. PSNR results of 4 datasets used in~\cite{Liu18TPAMI}, in each cell, top-left: \texttt{Kobe}, top-right: \texttt{Traffic}, middle-left: \texttt{Runner}, middle-right: \texttt{Drop}, bottom: average. The bold number denotes the highest PSNR (based on the 0.001 precision) for each projection algorithm per video dataset. The \textcolor{red}{red} number denotes the highest PSNR for each dataset across all the algorithms. {\em Italian} denotes the highest average PSNR for each row and the \textcolor{blue}{\em blue Italian} one is the highest average PSNR across all algorithms.}
	\resizebox{.48\textwidth}{!}				
	{
		\centering
		\begin{tabular}{|c||c|c|c||c|c|c|c|}
			\hline  & \multicolumn{3}{c||}{ATV} &
			\multicolumn{2}{c|}{ITV2D} &
			\multicolumn{2}{c|}{ITV3D} \\ 
			\cline{2-8} & Clip & Cham& FGP& Cham& FGP& Cham& FGP\\
			\hline FISTA  &\makecell{22.49, 18.80\\25.61, 29.40\\24.07}&\makecell{22.75, 18.80\\25.85, 29.62\\24.25}&\makecell{24.50, 19.97\\{27.82}, {32.13}\\26.11}& \makecell{23.11, 19.14\\26.50, 30.33\\24.77}&\makecell{{\bf 24.50}, 19.97\\{27.82}, 32.13\\26.11}&\makecell{23.29, 19.29\\26.47, 30.85\\24.97}&\makecell{{24.50}, 20.06\\{\bf 27.84}, {\bf 32.14} \\{\em 26.13} }\\
			\hline TwIST  &\makecell{25.38, 20.44\\28.12, 32.79\\26.68}&\makecell{25.47, 20.34\\28.24, 32.72\\26.69}&\makecell{25.83, 20.57\\{\bf 28.89}, {\bf 33.56}\\{27.21}}& \makecell{25.50, 20.37\\28.62, 32.97\\26.86}&\makecell{25.83, 20.56\\{28.89}, {33.56}\\27.21}&\makecell{24.98, 20.39\\28.02, 31.65\\26.26}&\makecell{25.78, 20.75\\28.86, 33.53\\{\em 27.23}}\\
			\hline GAP &\makecell{26.71, 20.75\\28.81, \textcolor{red}{\bf 33.97}\\\textcolor{blue}{\em 27.56}}&\makecell{\textcolor{red}{\bf 26.72}, 20.64\\28.91, 33.83\\  27.53}&\makecell{26.17, 20.67\\29.13, 33.91\\27.47}&\makecell{26.28, 20.53\\\textcolor{red}{\bf 29.13}, 33.74\\27.42}&\makecell{26.17, 20.65\\29.12, {33.91}\\27.46}&\makecell{25.38, 20.55\\28.33, 32.07\\26.58}&\makecell{26.10, \textcolor{red}{\bf 20.85}\\29.08, 33.87\\27.48}\\
			\hline ADMM &\makecell{25.88, 20.42\\28.61, 33.39\\27.08}&\makecell{26.05, 20.52\\28.58, 33.27\\27.10}&\makecell{26.00, 20.62\\{\bf 29.01}, {\bf 33.74}\\27.34}& \makecell{25.87, 20.44\\28.87, 33.36\\27.14}&\makecell{25.99, 20.61\\29.01, 33.74\\27.34}&\makecell{25.18, 20.47\\28.18, 31.94\\26.44}&\makecell{25.94, {\bf 20.80}\\28.98, 33.71\\{\em 27.36}}\\
			\hline
		\end{tabular}
	}
	\label{Tab:Summary}
	\vspace{-4mm}
\end{table}

Algorithm~\ref{Algo:GAP-ATV} gives the full algorithm of GAP with ATV using different denoising algorithms. It is easy to replace GAP with ADMM/TwIST/FISTA and replace ATV with ITV.
We thus achieve the various compositions of frameworks for SCI reconstruction with different TV denoising algorithms summarized in Table~\ref{Tab:Summary}.

\section{Experimental Results}
	\vspace{-3mm}
Now, we apply various TV algorithms and projection frameworks to video SCI on both simulation and real data.

\noindent{\bf Simulation:}
We used four datasets, \ie, \texttt{Kobe, Traffic, Runner, Drop} in~\cite{Liu18TPAMI}, where $B=8$ video frames are compressed into a single measurement and the same sensing matrix is used.
The results are summarized in Table~\ref{Tab:Summary}. It can be observed that, in general, the ITV3D algorithm works well in all scenarios. On average, the best result is obtained by GAP-ATV-Clip, though the gains over other approaches are very limited.
For each dataset, GAP provides the best result. Due to space limit, we did not show the reconstructed video frames here.
We also notice that only (In-Iter) 2 iterations can give good results of FPG while other TV solvers need 5 iterations.

\begin{figure}[htbp]	
	\vspace{-2mm}
	\centering
	{\includegraphics[width= 0.235\textwidth]{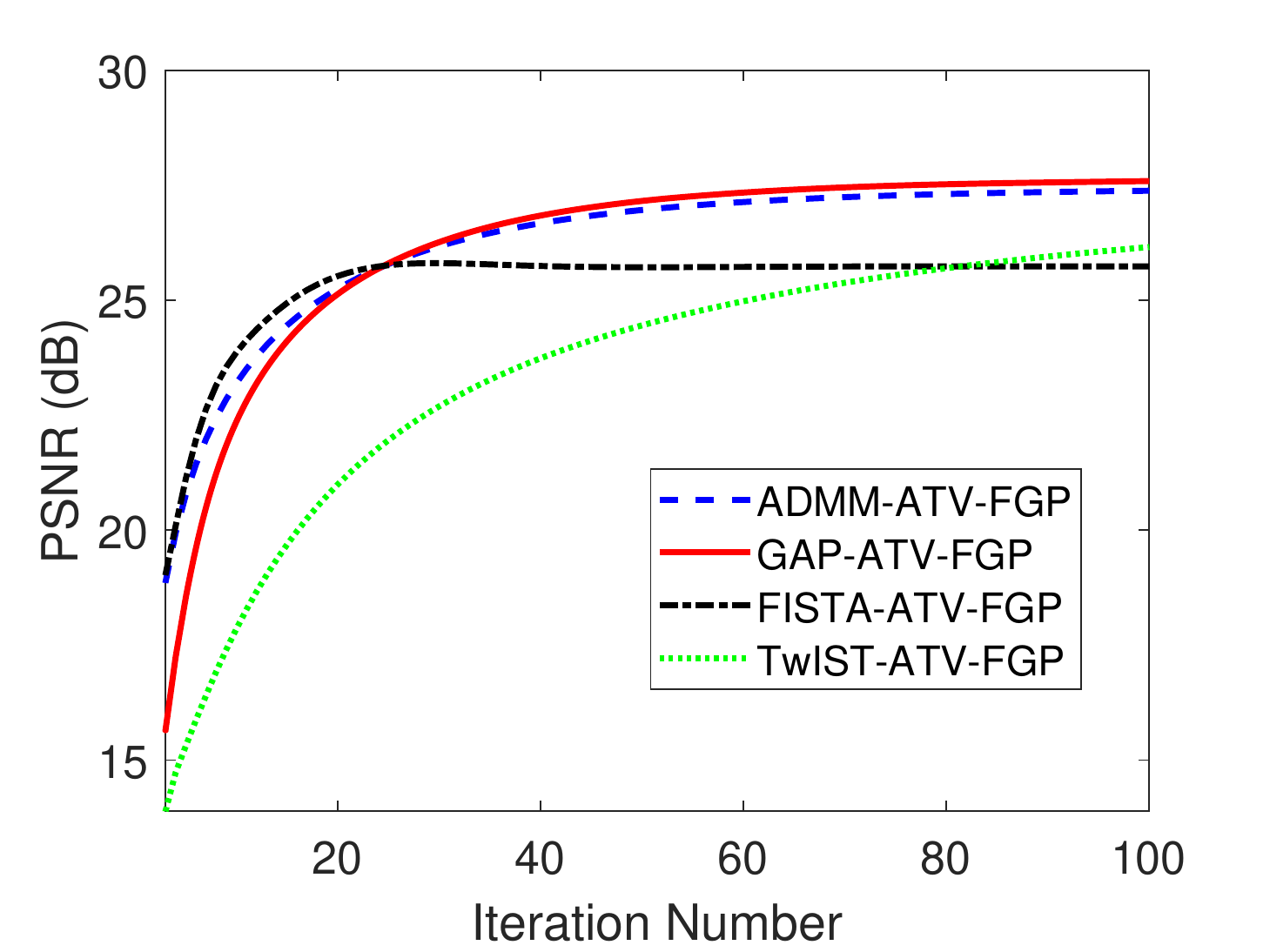}}
	{\includegraphics[width= 0.235\textwidth]{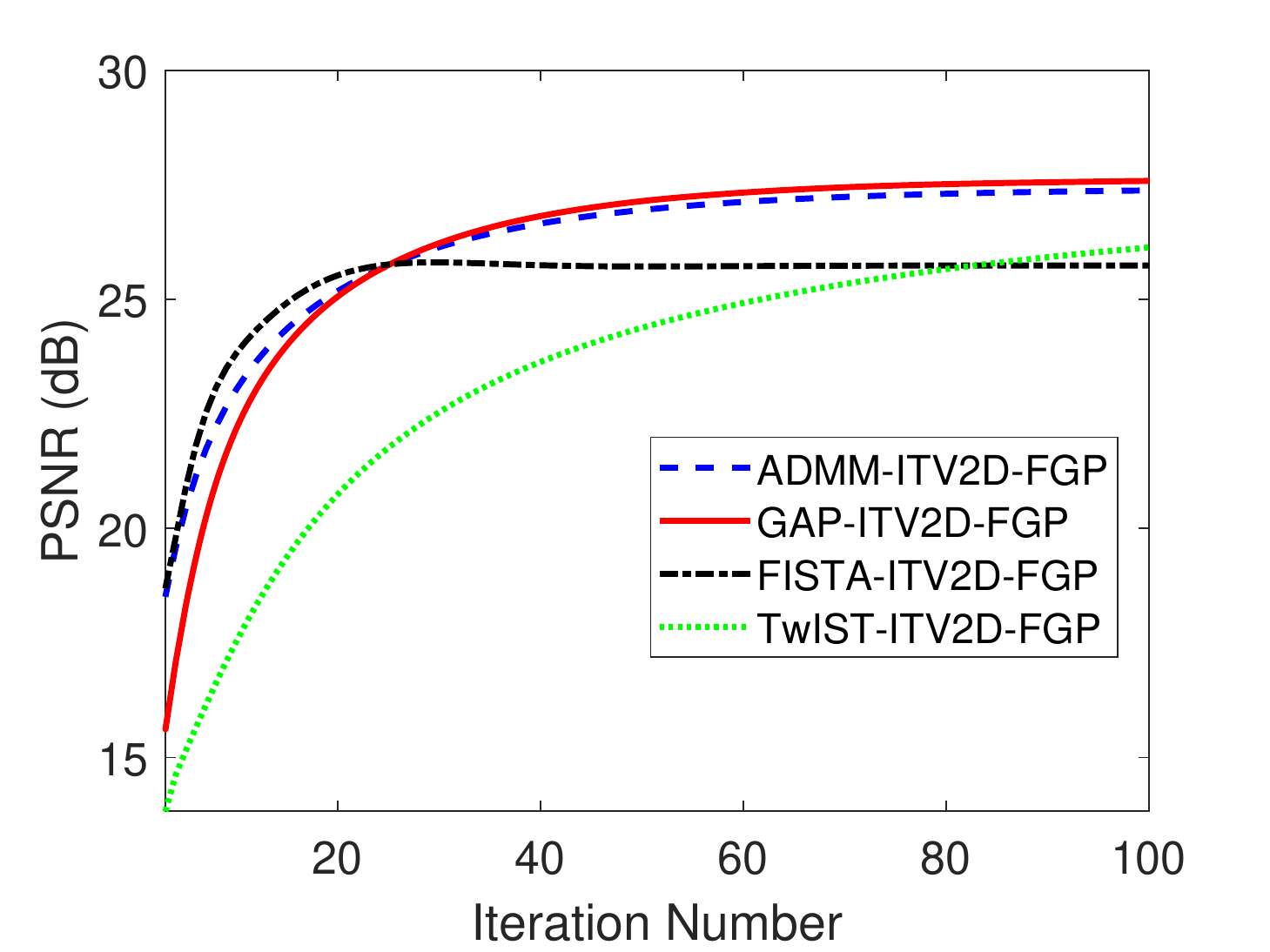}}
	\vspace{-3mm}
	\caption{\small PSNR vs. Iteration Number for different algorithms.}
	\label{fig:PSNR_iter}
	\vspace{-5mm}
\end{figure}

Another metric to compare different algorithms is the speed. As a good candidate for each row, we select ATV-FGP and ITV2D-FGP as the TV denoising algorithm and the first measurement in \texttt{Drop} is employed to show the reconstruction PSNR vs. iteration number in Fig.~\ref{fig:PSNR_iter}. It can be observed that TwIST always converges slowest and it usually needs 500 iterations to get a good result. FISTA converges fastest but cannot lead to good results while GAP and ADMM converge similarly to FISTA and GAP leads to the best results in about 60 iterations. 

\begin{figure}[htbp]	
	\vspace{-2mm}
	\centering
	{\includegraphics[width= 0.5\textwidth]{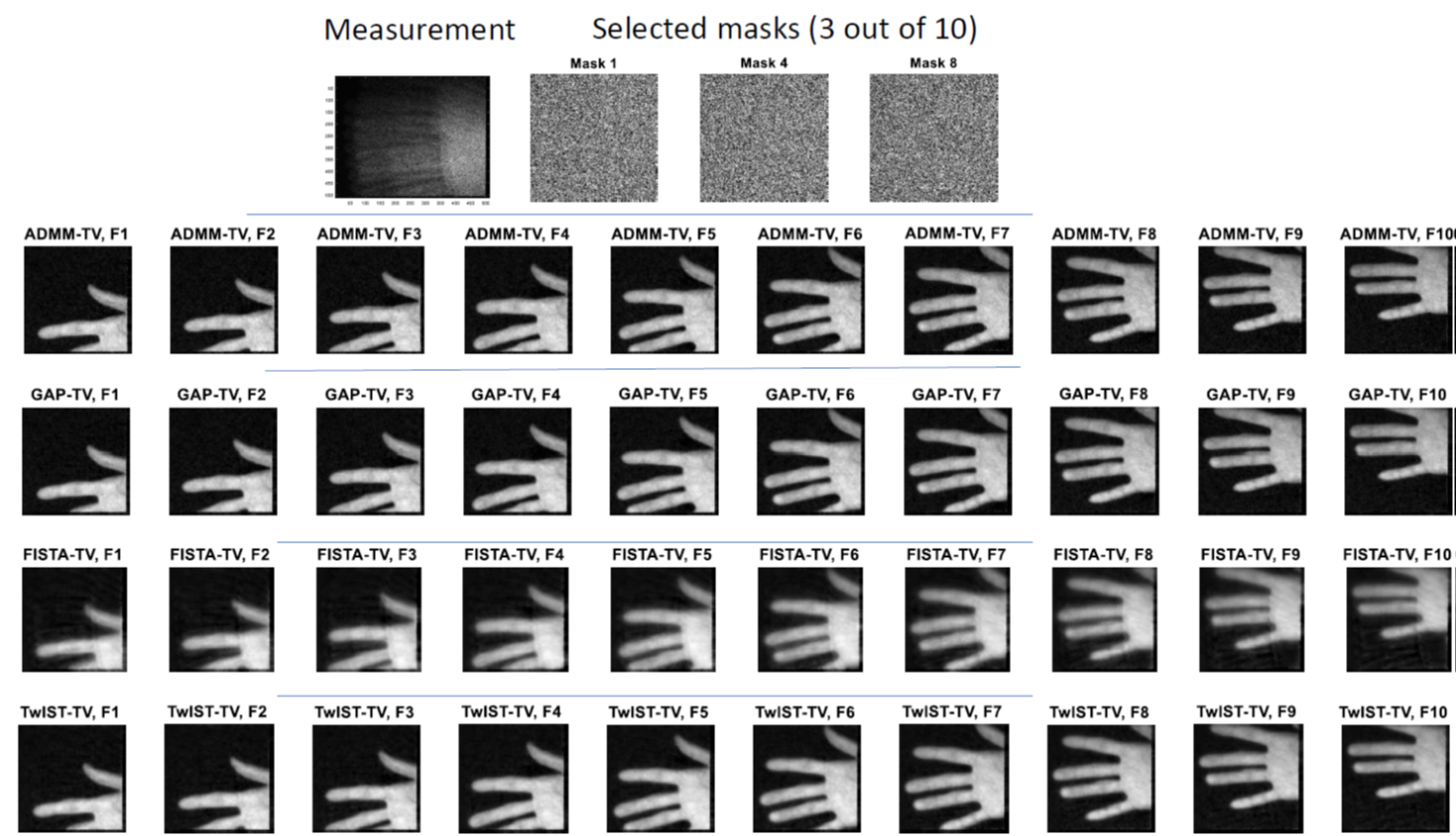}}
	\vspace{-8mm}
	\caption{\small Real data results. A hand is moving in front of a SCI camera and 10 frames are reconstructed from a snapshot measurement (top-left). 3 masks out of 10 are shown in the first row.}
	\label{fig:Hand_real}
	\vspace{-5mm}
\end{figure}

\noindent{\bf Real Data:}
We now test different methods on the real data captured by our video SCI camera. Our camera is similar to the design of~\cite{Hitomi11ICCV,Sun17OE}, which used a digital micromirror
device (DMD) to modulate the high-speed scene. In total, $B=10$ frames are modulated and compressed to a snapshot measurement (Fig.~\ref{fig:Hand_real} top-left) with a spatial resolution of $512\times512$. A hand is moving fast in front of the camera being a high-speed scene. 

ITV3D-FGP is used for the TV denoising and we run the algorithms for 150 iterations, which takes about 3 minutes on an Intel i7 CPU laptop with 32G memory. 
It can be seen that the all algorithms can reconstruction the motion clearly from the smashed (blurry) single measurement.
FISTA results suffers from blurry and the other 3 results look similar while GAP seems providing the best one. 

\begin{figure}[!htbp]	
	\vspace{-2mm}
	\centering
	{\includegraphics[width= 0.52\textwidth]{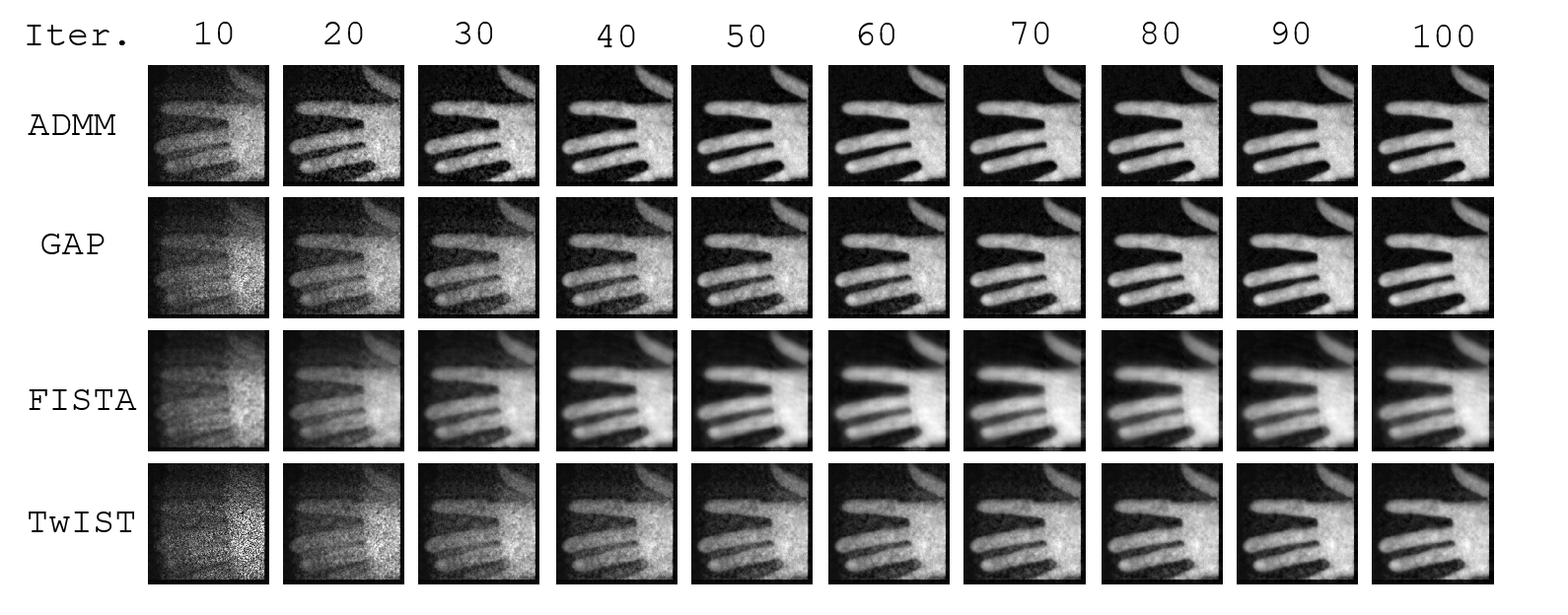}}
	\vspace{-10mm}
	\caption{\small Results by plotting Frame 6 every 10 iterations. }
	\label{fig:Hand_plot10}
	\vspace{-8mm}
\end{figure}

We again test the speed of different algorithms but this time by visualizing the results for every 10 iterations in Fig.~\ref{fig:Hand_plot10}.
It can be seen that FISTA can provide decent results in 40 iterations. However, it does not get improved with more iterations. This is similar to the simulation results.
ADMM is the second efficient one to give good results in 50 iterations and they are getting better with more iterations. TwIST is slowest and it needs over 100 iterations to get good results.

\section{Conclusions and Future Work}
We have investigated diverse total variation algorithms under different projection frameworks for video snapshot compressive imaging. GAP and ADMM are recommended for decent results while FISTA is a choice with limited running time. Regrading the total variation solver, FGP is recommended for different cases because it is faster. 
We are working on using these algorithms to initialize complicated algorithms~\cite{Yang14GMM,Yang14GMMonline,Miao19ICCV} like DeSCI to get better results.
Convergence results of these algorithms will also be derived.
We also found that a recent research line of Plug-and-Play framework~\cite{Yuan2020_CVPR_PnP} is interesting to be investigated for SCI reconstruction.

\footnotesize
\bibliographystyle{IEEEtran}

\begin{thebibliography}{10}
	\providecommand{\url}[1]{#1}
	\csname url@samestyle\endcsname
	\providecommand{\newblock}{\relax}
	\providecommand{\bibinfo}[2]{#2}
	\providecommand{\BIBentrySTDinterwordspacing}{\spaceskip=0pt\relax}
	\providecommand{\BIBentryALTinterwordstretchfactor}{4}
	\providecommand{\BIBentryALTinterwordspacing}{\spaceskip=\fontdimen2\font plus
		\BIBentryALTinterwordstretchfactor\fontdimen3\font minus
		\fontdimen4\font\relax}
	\providecommand{\BIBforeignlanguage}[2]{{%
			\expandafter\ifx\csname l@#1\endcsname\relax
			\typeout{** WARNING: IEEEtran.bst: No hyphenation pattern has been}%
			\typeout{** loaded for the language `#1'. Using the pattern for}%
			\typeout{** the default language instead.}%
			\else
			\language=\csname l@#1\endcsname
			\fi
			#2}}
	\providecommand{\BIBdecl}{\relax}
	\BIBdecl
	
	\bibitem{Liu18TPAMI}
	Y.~Liu, X.~Yuan, J.~Suo, D.~Brady, and Q.~Dai, ``Rank minimization for snapshot
	compressive imaging,'' \emph{IEEE Transactions on Pattern Analysis and
		Machine Intelligence}, pp. 1--1, 2018.
	
	\bibitem{Hitomi11ICCV}
	Y.~Hitomi, J.~Gu, M.~Gupta, T.~Mitsunaga, and S.~K. Nayar, ``Video from a
	single coded exposure photograph using a learned over-complete dictionary,''
	in \emph{IEEE International Conference on Computer Vision (ICCV)}, 2011.
	
	\bibitem{Reddy11CVPR}
	D.~Reddy, A.~Veeraraghavan, and R.~Chellappa, ``{\rm P2C2}: Programmable pixel
	compressive camera for high speed imaging,'' \emph{IEEE Computer Vision and
		Pattern Recognition (CVPR)}, 2011.
	
	\bibitem{Patrick13OE}
	P.~Llull, X.~Liao, X.~Yuan, J.~Yang, D.~Kittle, L.~Carin, G.~Sapiro, and D.~J.
	Brady, ``Coded aperture compressive temporal imaging,'' \emph{Optics
		Express}, pp. 698--706, 2013.
	
	\bibitem{Yuan14CVPR}
	X.~Yuan, P.~Llull, X.~Liao, J.~Yang, G.~Sapiro, D.~J. Brady, and L.~Carin,
	``Low-cost compressive sensing for color video and depth,'' in \emph{IEEE
		Conference on Computer Vision and Pattern Recognition (CVPR)}, 2014.
	
	\bibitem{Sun16OE}
	Y.~Sun, X.~Yuan, and S.~Pang, ``High-speed compressive range imaging based on
	active illumination,'' \emph{Opt. Express}, vol.~24, no.~20, pp.
	22\,836--22\,846, Oct 2016.
	
	\bibitem{Sun17OE}
	------, ``Compressive high-speed stereo imaging,'' \emph{Opt. Express},
	vol.~25, no.~15, pp. 18\,182--18\,190, July 2017.
	
	\bibitem{Yuan17AO}
	X.~Yuan, Y.~Sun, and S.~Pang, ``Compressive video sensing with side
	information,'' \emph{Appl. Opt.}, vol.~56, no.~10, pp. 2697--2704, 2017.
	
	\bibitem{Yuan&Pang16_ICIP}
	X.~Yuan and S.~Pang, ``Compressive video microscope via structured
	illumination,'' in \emph{2016 IEEE International Conference on Image
		Processing (ICIP)}, Sept 2016, pp. 1589--1593.
	
	\bibitem{Yuan13ICIP}
	X.~Yuan, J.~Yang, X.~Liao, P.~Llull, G.~Sapiro, D.~J. Brady, and L.~Carin,
	``Adaptive temporal compressive sensing for video,'' \emph{IEEE International
		Conference on Image Processing}, pp. 1--4, 2013.
	
	\bibitem{Gehm07}
	M.~E. Gehm, R.~John, D.~J. Brady, R.~M. Willett, and T.~J. Schulz,
	``Single-shot compressive spectral imaging with a dual-disperser
	architecture,'' \emph{Optics Express}, vol.~15, pp. 14\,013--14\,027, 2007.
	
	\bibitem{Wagadarikar08CASSI}
	A.~Wagadarikar, R.~John, R.~Willett, and D.~J. Brady, ``Single disperser design
	for coded aperture snapshot spectral imaging,'' \emph{Applied Optics},
	vol.~47, no.~10, pp. B44--B51, 2008.
	
	\bibitem{Wagadarikar09CASSI}
	A.~Wagadarikar, N.~Pitsianis, X.~Sun, and D.~Brady, ``Video rate spectral
	imaging using a coded aperture snapshot spectral imager,'' \emph{Optics
		Express}, vol.~17, no.~8, pp. 6368--6388, 2009.
	
	\bibitem{Yuan15JSTSP}
	X.~Yuan, T.-H. Tsai, R.~Zhu, P.~Llull, D.~J. Brady, and L.~Carin, ``Compressive
	hyperspectral imaging with side information,'' \emph{IEEE Journal of Selected
		Topics in Signal Processing}, vol.~9, no.~6, pp. 964--976, September 2015.
	
	\bibitem{Cao16SPM}
	X.~Cao, T.~Yue, X.~Lin, S.~Lin, X.~Yuan, Q.~Dai, L.~Carin, and D.~J. Brady,
	``Computational snapshot multispectral cameras: Toward dynamic capture of the
	spectral world,'' \emph{IEEE Signal Processing Magazine}, vol.~33, no.~5, pp.
	95--108, Sept 2016.
	
	\bibitem{Donoho06ITT}
	D.~L. Donoho, ``Compressed sensing,'' \emph{IEEE Transactions on Information
		Theory}, vol.~52, no.~4, pp. 1289--1306, April 2006.
	
	\bibitem{Candes06ITT}
	E.~J. Cand\`{e}s, J.~Romberg, and T.~Tao, ``Robust uncertainty principles:
	Exact signal reconstruction from highly incomplete frequency information,''
	\emph{IEEE Transactions on Information Theory}, vol.~52, no.~2, pp. 489--509,
	February 2006.
	
	\bibitem{Jalali19TIT}
	S.~Jalali and X.~Yuan, ``Snapshot compressed sensing: performance bounds and
	algorithms,'' \emph{IEEE Transactions on Information Theory}, 2019.
	
	\bibitem{Jalali18ISIT}
	------, ``Compressive imaging via one-shot measurements,'' in \emph{2018 IEEE
		International Symposium on Information Theory (ISIT)}, June 2018, pp.
	416--420.
	
	\bibitem{Beck09IST}
	A.~Beck and M.~Teboulle, ``A fast iterative shrinkage-thresholding algorithm
	for linear inverse problems,'' \emph{SIAM J. Img. Sci.}, vol.~2, no.~1, pp.
	183--202, Mar. 2009.
	
	\bibitem{Bioucas-Dias2007TwIST}
	J.~Bioucas-Dias and M.~Figueiredo, ``A new {TwIST}: Two-step iterative
	shrinkage/thresholding algorithms for image restoration,'' \emph{IEEE
		Transactions on Image Processing}, vol.~16, no.~12, pp. 2992--3004, December
	2007.
	
	\bibitem{Liao14GAP}
	X.~Liao, H.~Li, and L.~Carin, ``Generalized alternating projection for
	weighted-$\ell_{2,1}$ minimization with applications to model-based
	compressive sensing,'' \emph{SIAM Journal on Imaging Sciences}, vol.~7,
	no.~2, pp. 797–--823, 2014.
	
	\bibitem{ADMM2011Boyd}
	S.~Boyd, N.~Parikh, E.~Chu, B.~Peleato, and J.~Eckstein, ``Distributed
	optimization and statistical learning via the alternating direction method of
	multipliers,'' \emph{Found. Trends Mach. Learn.}, vol.~3, no.~1, pp. 1--122,
	January 2011.
	
	\bibitem{Yuan16ICIP_GAP}
	X.~Yuan, ``Generalized alternating projection based total variation
	minimization for compressive sensing,'' in \emph{2016 IEEE International
		Conference on Image Processing (ICIP)}, Sept 2016, pp. 2539--2543.
	
	\bibitem{Selesnick09Clip}
	I.~Selesnick and I.~Bayram, ``Total variation filtering,'' \emph{Connexions},
	2009.
	
	\bibitem{Beck09TV}
	A.~Beck and M.~Teboulle, ``Fast gradient-based algorithms for constrained total
	variation image denoising and deblurring problems,'' \emph{IEEE Transactions
		on Image Processing}, vol.~18, no.~11, pp. 2419--2434, Nov. 2009.
	
	\bibitem{Chambolle2005TVM}
	A.~Chambolle, ``Total variation minimization and a class of binary mrf
	models.''
	
	\bibitem{Zhu2010}
	M.~Zhu, S.~J. Wright, and T.~F. Chan, ``Duality-based algorithms
	for total-variation-regularized image restoration,'' \emph{Computational
		Optimization and Applications}, vol.~47, no.~3, pp. 377--400, Nov 2010.
	
	\bibitem{Chambolle2004ATV}
	A.~Chambolle, ``An algorithm for total variation minimization and
	applications,'' \emph{J. Math. Imaging Vis.}, vol.~20, no. 1-2, pp. 89--97,
	Jan. 2004.
	
	\bibitem{Chan_2008dualTV}
	X.~Bresson and T.~F. Chan, ``Fast dual minimization of the vectorial total
	variation norm and applications to color image processing,'' \emph{Inverse
		Problems and Imaging}, vol.~2, p. 455, 2008.
	
	\bibitem{Yang14GMM}
	J.~Yang, X.~Yuan, X.~Liao, P.~Llull, G.~Sapiro, D.~J. Brady, and L.~Carin,
	``Video compressive sensing using {G}aussian mixture models,'' \emph{IEEE
		Transaction on Image Processing}, vol.~23, no.~11, pp. 4863--4878, November
	2014.
	
	\bibitem{Yang14GMMonline}
	J.~Yang, X.~Liao, X.~Yuan, P.~Llull, D.~J. Brady, G.~Sapiro, and L.~Carin,
	``Compressive sensing by learning a {G}aussian mixture model from
	measurements,'' \emph{IEEE Transaction on Image Processing}, vol.~24, no.~1,
	pp. 106--119, January 2015.
	
	\bibitem{Miao19ICCV}
	X.~Miao, X.~Yuan, Y.~Pu, and V.~Athitsos, ``$\lambda$-net: Reconstruct
	hyperspectral images from a snapshot measurement,'' in \emph{IEEE/CVF
		Conference on Computer Vision (ICCV)}, 2019.
	
	\bibitem{Yuan2020_CVPR_PnP}
	X.~{Yuan}, Y.~{Liu}, J.~{Suo}, and Q.~{Dai}, ``Plug-and-play algorithms for
	large-scale snapshot compressive imaging,'' in \emph{CVPR}, June 2020.
	
\end{thebibliography}

\end{document}